\documentstyle[epsf,11pt]{article}

\begin{document}

\vspace*{-1.8cm}

\begin{flushright}
{\large\bf LAL 06-145}\\
\vspace*{0.1cm}
{\large October 2006}
\end{flushright}
\vspace*{1,3cm}

\begin{center}
{\LARGE\bf The Impact of BeamCal Performance at\\
\vspace*{0,1cm}

 Different ILC Beam Parameters and \\
\vspace*{0,3cm}

Crossing Angles on $\tilde{\tau}$~searches}
\end{center}
\vspace*{1cm}

\begin{center}
{\large\bf P. Bambade$^a$,} \\
\vspace*{0,2cm}
$^a$ {\bf Laboratoire de l'Acc\'{e}l\'{e}rateur Lin\'{e}aire}\\ 
CNRS/IN2P3 and Universit\'{e} Paris Sud 11, B\^{a}t. 200, B.P. 34, F-91898 Orsay, Cedex, France\\
\vspace*{0,7cm}

{\large\bf V. Drugakov$^b$\footnote{Talk given at the Linear Collider Workshop ``{\it\bf LCWS06}'',9-13 March 2006, I.I.Sc Bangalore, India}}\\
\vspace*{0,2cm}
$^b$ {\bf NC PHEP}, M.Bogdanovich 153, 220040 Minsk, Belarus\\
\vspace*{0,7cm}

{\large\bf W. Lohmann$^c$}\\
\vspace*{0,2cm}
$^c$ {\bf DESY}, Platanenallee 6, D-15738 Zeuthen, Germany
\end{center}
\vspace*{1cm}
\begin{abstract}
The ILC accelerator parameters and detector concepts are still under discussion in the
world-wide community. As will be shown, the performance of the BeamCal, the
calorimeter in the very forward area of the ILC detector, is very
sensitive to the beam parameter and crossing angle choices. We propose here
BeamCal designs for a small (0 or 2~mrad) and large (20~mrad) crossing angles
and report about the veto performance study done. As illustration, the influence of several proposed beam parameter sets and crossing-angles on the signal to background ratio in the stau search is estimated for a particular realization of the super-symmetric model.
\end{abstract}
\vspace*{0,5cm}

\section{Introduction}
\begin{figure}h]
  \begin{tabular}{lr}
    \epsfxsize=0.47\textwidth
    \epsfbox{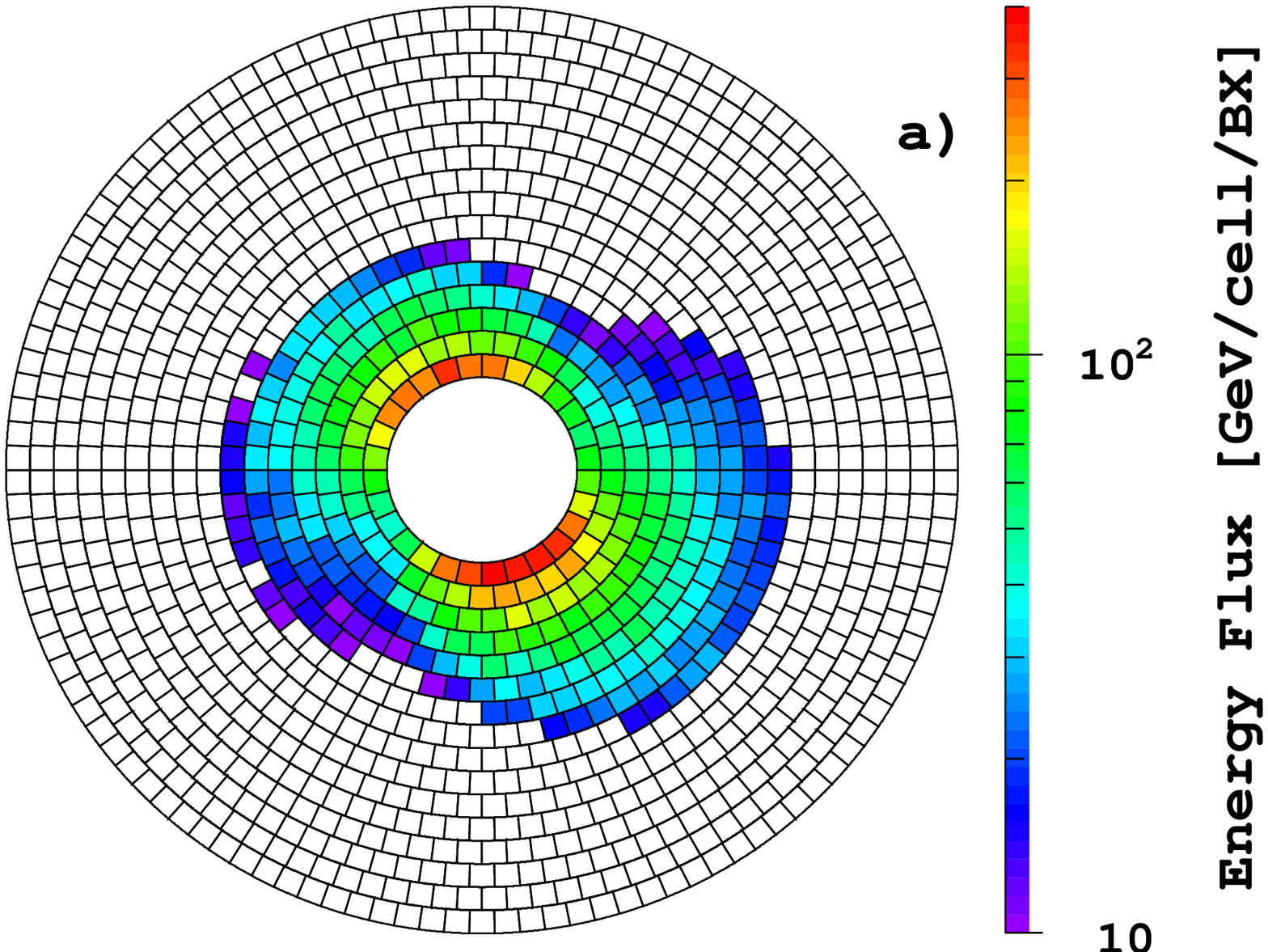}&
    \epsfxsize=0.47\textwidth
    \epsfbox{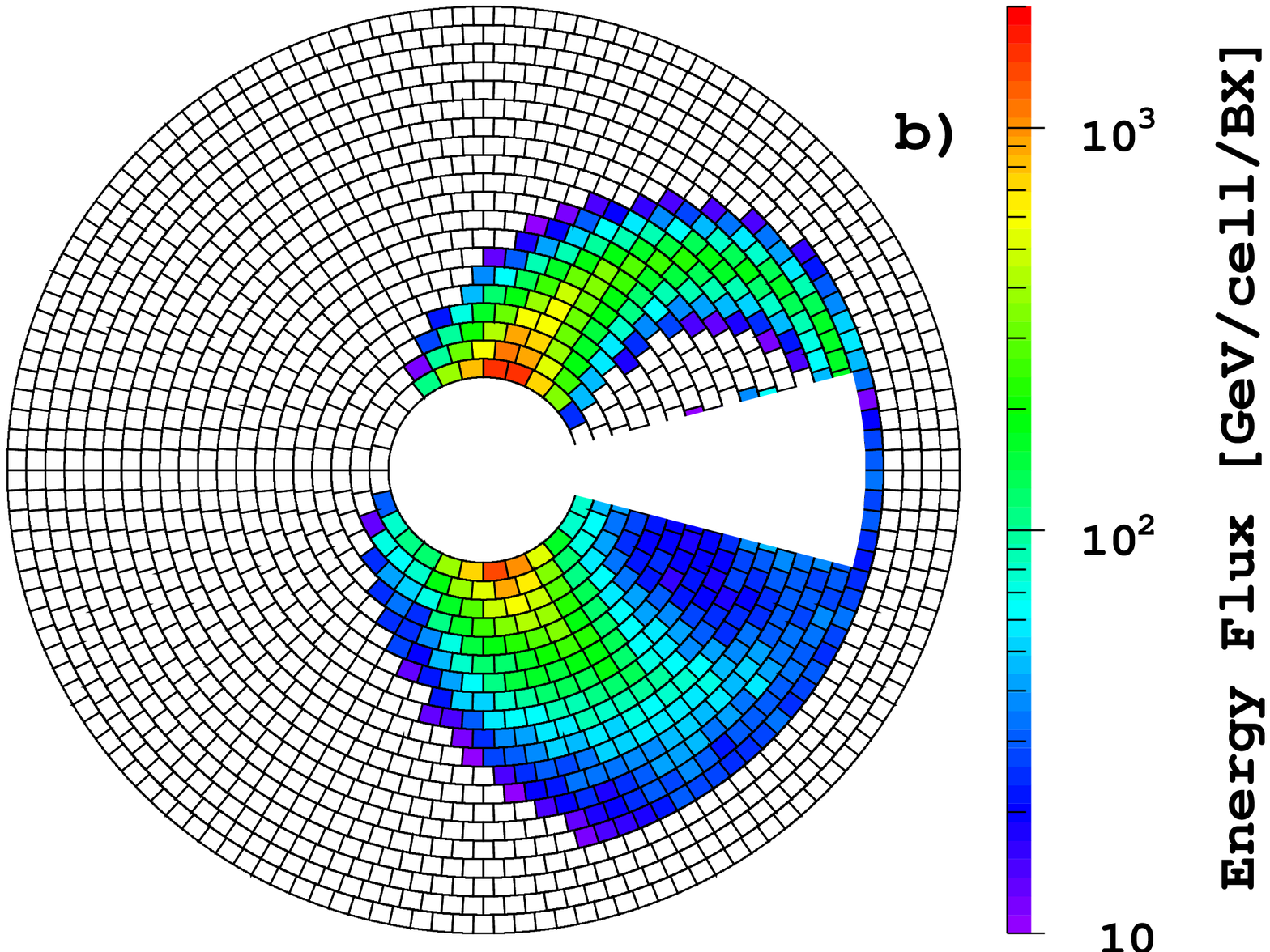}
  \end{tabular}
  \caption{\it The energy density of beamstrahlung remnants per bunch crossing
    as a function of position in the $r-\varphi$ plane at the a) 2~mrad
    and b) 20~mrad with DID field crossing angles.}
  \label{fig:EfluxX}
\end{figure}

The TESLA machine parameters were chosen to achieve a high peak luminosity
with only little room for operational optimization. As a more flexible
approach~\cite{raubenheimer} defines a number of different machine
configurations achieving similar peak luminosities. We consider here the impact
of these schemes on the pair depositions in the BeamCal. The pairs, stemming
from the beamstrahlung photon conversions, deposit several TeV of energy in the
BeamCal (see Fig.~\ref{fig:EfluxX}) with large local energy density
fluctuations from bunch to bunch. Identification of single electrons on top of
these depositions is challenging at the inner part of the BeamCal even at the
highest electron energies~\cite{PRC}.

For a 2~mrad crossing angle the depositions of the pairs in the BeamCal are
very similar to the ones of the head-on scheme (Fig.~\ref{fig:EfluxX}a). The
only change required for the BeamCal is a slightly larger inner radius.

The 20 mrad crossing angle geometry is proposed with several possible arrangements for the magnetic field inside the detector.
In the DID\footnote{Detector integrated dipole}~\cite{DID} field configuration the
magnetic field is directed along the incoming beam lines with a kink at the
transverse plane containing the IP. Hence the incoming beams will not emit synchrotron radiation and their spins will not precess before they collide. However, the amount of beamstrahlung deposited in the BeamCal rises considerably (Fig.~\ref{fig:EfluxX}b), causing
also higher background in the tracking detectors due to
backscattering. Conversely, if the magnetic field is directed along outgoing
beam lines with a kink at the IP plane (a configuration referred to as
anti-DID), the depositions on the BeamCal and background in the central detector are very similar to the head-on case.

The BeamCal is an important tool to identify two photon events by detecting either
electrons or positrons with an energy near the beam energy. Two photon events
constitute the most serious background for many search channels which are
characterized by missing energy and missing momentum. In most cases lepton
pairs produced in photon-photon processes have significantly different
topology and kinematics in comparison to the search channel and can be
rejected by simple cuts. However, since the two photon cross-section is
typically several orders of magnitude larger, events in the tails of the
kinematic distributions become important. 

The electron veto performance, obtained from simulations, is used to estimate
the suppression of the two-photon background in the different ILC schemes. The
search for a $\tilde{\tau}$~super-symmetric particle is taken as a benchmark
process, following the approach described in~\cite{bambade}. In the particular realization of the super-symmetric model, which is consider here (point 3 in the list of SUSY benchmark points for the ILC
detector~\cite{PhysBench}), the $\tilde{\tau}$'s are the second lightest
super-symmetric particles which are pair-produced in e$^+$e$^-$ annihilation and decay into lighter
neutralinos, which escape undetected, and regular $\tau$'s. In the context of this
model, the $\tilde{\tau}$'s and neutralinos could combine to provide a
plausible, quantitative explanation for the amount of dark matter in the
universe. The amount is directly related to the mass difference between
$\tilde{\tau}$ and neutralino and is assumed here to be equal 5~GeV.

\section{Simulation and Results}

Single electrons and beamstrahlung pairs were simulated for 4 proposed
accelerator parameter sets (Tab.~\ref{tab:BeamParamList}) at 
zero crossing angle and for the Nominal set at 20 mrad crossing angle with the DID magnetic field configuration. Beamstrahlung was generated using GUINEA-PIG~\cite{gpig}. The detector was simulated in GEANT4~\cite{geant4}. 

\begin{table}[h]
\vspace*{-0,2cm}
  \begin{center}
    \begin{tabular}{|l|c|c|c|c|}
      \hline
      \bf & Nominal & LowQ & LargeY & LowP \\
      \hline
      \hline
      Bunch charge [10$^{10}$] & 2 & 1 & 2 & 2 \\
      \hline
      Number of bunches & 2820 & 5640 & 2820 & 1330 \\
      \hline
      $\gamma\epsilon_{x}$/$\gamma\epsilon_{y}$ [10$^{-6}$ mrad] & 10
      / 0.04 & 10 /0.03 & 12 / 0.08 & 10 / 0.035 \\
      \hline
      $\beta_{x}$ / $\beta_{y}$ [mm] & 21 / 0.4  & 12 / 0.2 & 10 / 0.4 & 10
      / 0.2\\
      \hline
      $\sigma_{x}$ / $\sigma_{y}$ [nm] & 655 / 5.7  & 495 / 3.5 & 495 / 8.1 &
      452 / 3.8\\
      \hline
      $\sigma_{z}$ [$\mu$m] & 300 & 150 & 500 & 200 \\
      \hline
      Luminosity [10$^{34}$ cm$^{-2}$s$^{-1}$] & 2.03 & 2.01 & 2.00 & 2.05 \\
      \hline
    \end{tabular}
    \caption{\it Beam and IP parameters for various beam parameter configurations
    at $\sqrt{s}$~=~500~GeV.}
    \label{tab:BeamParamList}
  \end{center}
\vspace*{-0,2cm}
\end{table}

The BeamCal is located 370 cm from the interaction point. The inner radius is 1.5
cm for 0 mrad crossing angle and 2 cm for 20 mrad. The outer radius is 16.5 cm.
For the 20 mrad crossing angle area of 30 degrees in $r-\phi$ plane between
the beam pipes for the in and outgoing beams is assumed to be 100\% inefficient
for particle detection, in anticipation of the practical difficulties to
instrument this area.

Fine granularity is necessary to identify the depositions from high
energy electrons and photons on top of energy depositions from beamstrahlung
remnants. The simulated sampling calorimeter is longitudinally divided into 30
disks of tungsten, each 1~X$_0$ thick (3.5~mm) interleaved by diamond active layers
(0.5~mm). The sensitive planes are divided into pads with a size of about half
a Moli\`ere radius (5~mm) in both dimensions, as shown in Fig.~\ref{fig:EfluxX}.

The energy depositions from pairs and the single electrons depositions are
superimposed in the sensor pads. A reconstruction algorithm is applied to the
output. The reconstruction procedure is described in more details in \cite{ieee}.

\begin{figure}[h]
  \begin{tabular}{lr}
    \epsfxsize=0.57\textwidth
\hspace*{-0,7cm}
    \epsfbox{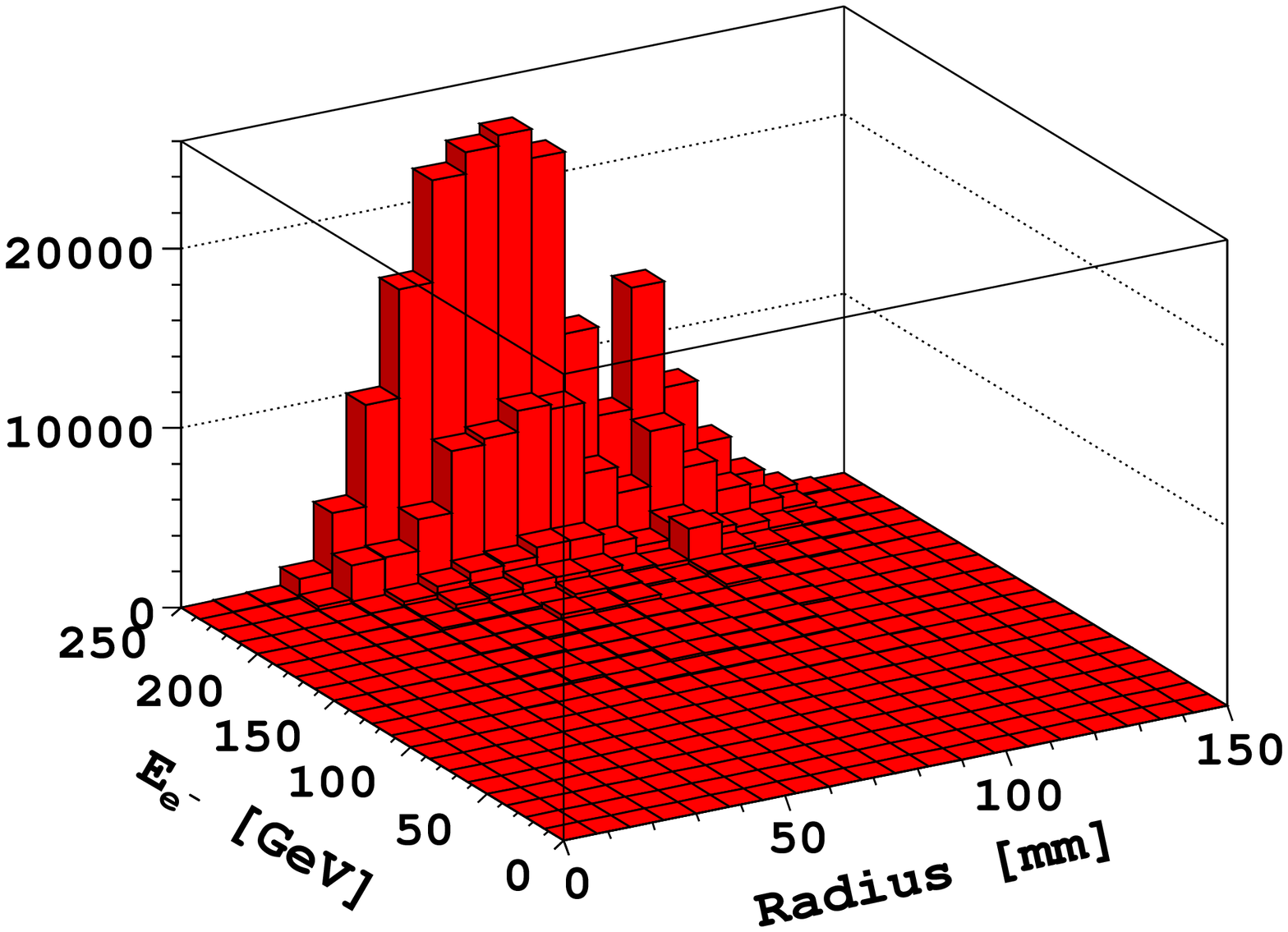}&
\hspace*{-0,5cm}
    \epsfxsize=0.47\textwidth
    \epsfbox{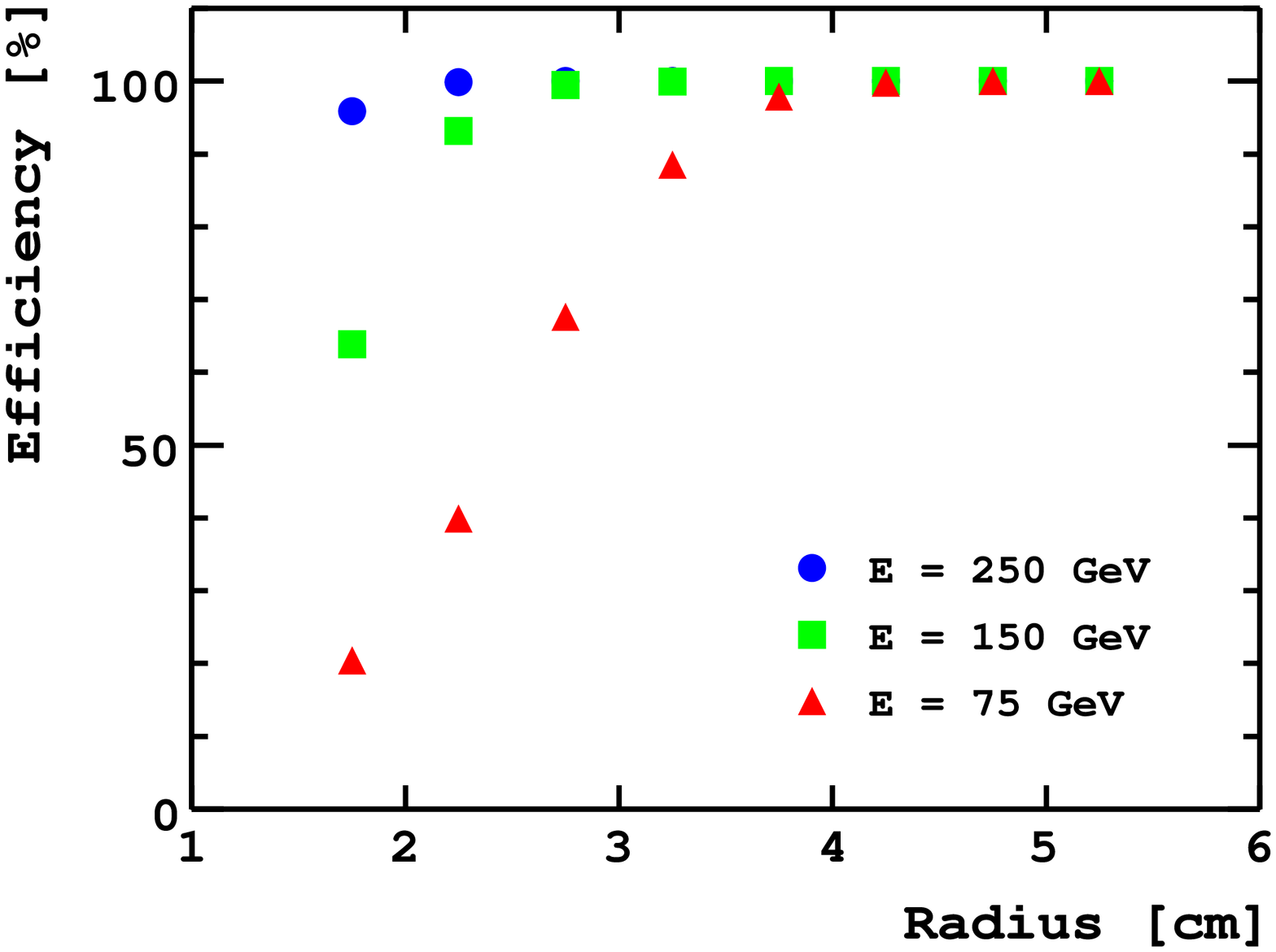}
  \end{tabular}
  \caption{\it Left: Electron energy and spatial distribution of the 2-photon
    background events passed all selection cuts except the BeamCal veto. Right: The efficiency to veto an electron of energy 75, 150, 250~GeV as a function of the radius in the BeamCal.}
  \label{fig:2photonbg}
\end{figure}

To evaluate the ability to suppress two-photon processes, we used the energy and polar angle distributions of the two-photon background events remaining in the
analysis described in~\cite{bambade} after application of all selection cuts other than the electron veto~\cite{zhang}. At this stage of the analysis, 20 stau events are left, while the number of surviving 2-photon background events is about 2.7$\times$10$^{5}$. Fig.~\ref{fig:2photonbg} shows the energy and spatial distributions of these electrons. Most of them have nearly the beam energy and hit the BeamCal
outside the area affected by pairs, though the distribution has tails down to the smallest angles and energies. It is important to notice that this distribution depends on the particular stau-neutralino mass difference considered. In this study it is 5 GeV, if it would be larger (smaller), for example the polar angle distribution would broaden (sharpen) and shift to larger (smaller) values.

The average efficiency to veto electrons is shown in Fig.~\ref{fig:2photonbg}
for several electron energies for head-on collisions and Nominal beam parameters. An electron of 250 GeV is vetoed even in regions with high
background with almost 100\% efficiency. The efficiency drops near the
innermost radius, partly due to shower leakage. Electrons of 75 GeV are
identified with high efficiency only at larger radii.

Depending on the cuts in the reconstruction algorithm, fake electron candidates can also be found. This can be either high energy particles from tails of the incoherent pair production process or background fluctuations which mimic the electron signal. In this study, the reconstruction algorithm was tuned  for a misidentification rate of 10\%.

For each beam parameter set in table~\ref{tab:BeamParamList} veto efficiencies
 are estimated from simulations in the instrumented area of the BeamCal (see figure~\ref{fig:EfluxX}). These efficiencies were used to determine the number of
remaining non-vetoed two-photon background events in the stau analysis, see
table~\ref{tab:BeamParRes}. Results are given for energy cuts of 50 and 75~GeV, showing that a relatively low energy cut of 50 GeV reduces considerably this
background contribution. For the chosen benchmark physics scenario the chances to see $\tilde{\tau}$~particles are very good for most of the accelerator designs, except the Low P scheme in which this remaining background completely dominates the selected event sample. By far the best situation is obtained for the Low Q scheme.

\begin{table}[h]
  \begin{center}
    \begin{tabular}{l|cc}
      Energy cut [GeV] & 75 & 50 \\
      \hline
      Nominal, 0~mrad & 45 & 5 \\
      LowQ, 0~mrad & 40 & 0.1 \\
      LargeY, 0~mrad & 50 & 9 \\
      LowP, 0~mrad & 364 & 321 \\
      Nominal, 20~mrad, DID & 396 & 349 \\
    \end{tabular}
    \caption{\it The number of un-vetoed background events. The number of $\tilde{\tau}$~events is 20.}
    \label{tab:BeamParRes}
  \end{center}
\end{table}

In the cases of 2 mrad or 20~mrad with the anti-DID field
configuration we expect the BeamCal performance to be similar to that in the head-on scheme, as the corresponding pairs deposition distributions are similar.
In case of 20~mrad crossing angle with a DID field configuration, we would have no chance to see $\tilde{\tau}$~production for this benchmark scenario. 

For the 20 mrad crossing angle geometry, an additional reduction in expected signal-to-background ratio arises independent of choosing a DID or anti-DID magnetic field configuration, because removing events with electrons missed in the larger un-instrumented part of the BeamCal (see figure~\ref{fig:EfluxX}) requires additional special selection cuts~\cite{bambade} and because of the increased fake veto rate from Bhabha processes with only a single electron seen~\cite{PRC}. Estimations have shown that these additional effects would amount to about 30-50\% in total for the present super-symmetric scenario, which could be compensated for with additional luminosity.

\section*{Acknowledgments}{We would like to thank Z. Zhang (LAL, Orsay) for
  providing the samples of the simulated two-photon background events in the
stau analysis and helpful discussions.}

\end{document}